# Building and Maintaining Halls of Fame over a Database

Foteini Alvanaki, Sebastian Michel, Aleksandar Stupar

*Saarland University*
*Saarbrücken, Germany*
{alvanaki|smichel|astupar}@mmci.uni-saarland.de

*Abstract*— Halls of Fame are fascinating constructs. They represent the elite of an often very large amount of entities—persons, companies, products, countries etc. Beyond their practical use as static rankings, changes to them are particularly interesting—for decision making processes, as input to common media or novel narrative science applications, or simply consumed by users. In this work, we aim at detecting events that can be characterized by changes to a Hall of Fame ranking in an automated way. We describe how the schema and data of a database can be used to generate Halls of Fame. In this database scenario, by Hall of Fame we refer to distinguished tuples; entities, whose characteristics set them apart from the majority. We define every Hall of Fame as one specific instance of an SQL query, such that a change in its result is considered a noteworthy event. Identified changes (i.e., events) are ranked using lexicographic tradeoffs over event and query properties and presented to users or fed in higher-level applications. We have implemented a full-fledged prototype system that uses either database triggers or a Java based middleware for event identification. We report on an experimental evaluation using a real-world dataset of basketball statistics.

## I. INTRODUCTION

The concept of rankings is ubiquitous; it exists in nearly all domains. Essentially, rankings allow focusing on a small subset of an exhaustively full list—usually a few top or bottom entries are of interest. Such small subsets represent the essence of the available data, worthwhile to look into. We refer to the top-k portion of such rankings as Halls of Fame—or lists or charts—and specifically address rankings of real-world entities, such as persons, companies, cities, or products. The common ground of such Halls of Fame is that they rank a certain subset of all entities (e.g., companies in Europe) with respect to a numeric attribute, for instance, by revenue in case of companies.

Changes to a Hall of Fame are often picked up immediately by the media—ideally broadcast right after the change occurred during a live transmission, e.g., on February 20th, 2012: *"Today, Dallas Mavericks's Dirk Nowitzki entered the top 20 of the NBA all-time scoring list."* The 23,335 points he achieved back then was one more than Robert Parish scored in his career, who held rank 20 before Nowitzki topped him. A single additional point scored made him enter the top-20 all-time list. Granted, this very event is easily predictable upfront, at least for an NBA fan and supporter of Nowitzki and his team. But the point is that there is a multitude of such situations, in all kind of domains.

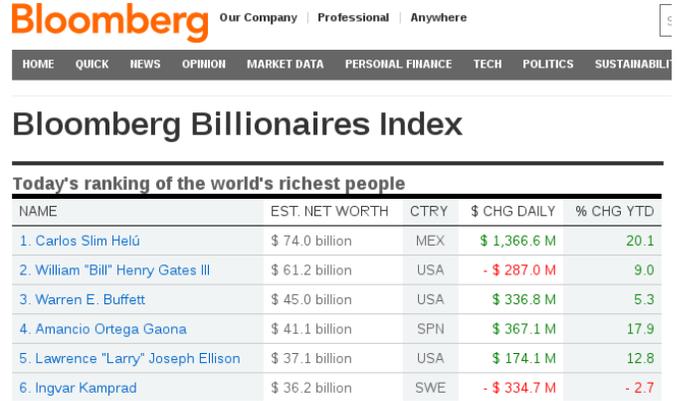

Fig. 1. Bloomberg's Billionaire Index, updated automatically, on a daily basis.

A salient showcase for the importance of computing and monitoring rankings is products: consumers are interested in rankings on the most effective flue vaccine, ranked by time to take effect, price, or number of adverse effects—likewise the entire pharmaceutical industry is interested, too. Changes to such rankings are particularly of interest: companies are interested in getting live feedback on the impact of their commercial campaigns by knowing how the latest products compete with the ones presented by competitors. This goes far beyond recent approaches on trend detection in Twitter that mainly look at changes to plain #hashtag popularities [1], [2]. A key challenge is to update such entity rankings automatically—like Bloomberg is doing with its Billionaires Index. This ranking—Figure 1 shows a subset—reports on a daily basis on the top-40 billionaires in the world. The update is done in an automated fashion; this is possible as Bloomberg knows how the wealth of the top billionaires is composed and affected by the stock market, current economic trends, etc. Having access to these individual numbers allows Bloomberg to update the Billionaires index.

Such a scenario is given as an illustrative example in Figure 2. The table *person* contains information like name, age, country of individuals, while *shareholder* describes the number of company stocks the person currently holds. The *stockmarket* relation consists of attributes that describe the current stock value of a company. Then, a Hall of Fame query that computes the top 10 richest people according to their stock

market shares is given as

**SELECT** p.p_name, **SUM**(s.s_shares * m.m_value)
**FROM** person p **JOIN** shareholder s **ON**
p.p_personid=s.s_personid **JOIN** market m
**ON** m.m_companyid=s.s_companyid
**GROUP BY** p.p_name
**ORDER BY SUM**(s.s_shares * m.m_value) **DESC**
**LIMIT 10**

This Hall of Fame ranking can be made more specific, for instance, through tailoring it to the country USA by adding the constraint country='USA' to the query. Additionally, or alternatively the ranking might use only technology companies for the wealth computation, maybe to investigate how exposed people are to fluctuations of the technology sector.

This illustrative example should highlight the core concepts of an automatically updatable entity ranking: data can be spread across tables, entities are ranked based on (aggregates of) numeric attributes, and the focus of the Hall of Fame is confined by constraints on attributes that reflect categories (such as country).

In this work, we show how to generate such queries based on the schema and data of a database, how to monitor changes to their results in the presence of database updates, and how to rank observed events in a way that appeals to users. To the best of our knowledge, this problem has not been addressed before.

*A. Contributions and Outline*

In this work, we make the following contributions:

- we define the concept of Halls of Fame of a database.
- we propose an intuitive framework for the generation of Halls of Fame, using combinations of categorical attributes, numerical attributes, and attributes that describe entities.
- we develop a scoring model to rank identified changes (i.e., events) in a meaningful way, using the static properties of a Hall of Fame, as well as dynamic aspects describing the change.
- we describe implementation details of two event detection approaches—based on database triggers and a Java middleware.
- we report on the results of a performance study showing the effectiveness and applicability of our approach.

This paper is organized as follows. Section II discusses related work. Section III introduces the data model and gives the problem statement. Section IV presents our framework to generate queries, each describing a Hall of Fame. Section V presents how the different facets covered by a Hall of Fame query can be described qualitatively and how these scores can be used for a final event ranking. Section VI discusses techniques to detect changes inside the Hall of Fame rankings. Section VII presents our experimental evaluation, followed by the conclusion in Section VIII.

## II. RELATED WORK

The work we present in this paper touches various research directions. The part of the paper introducing the concept of Hall of Fame rankings and how to generate them is mainly related to database mining. Once Hall of Fame rankings are determined, we deal with the problem of detecting changes to them, which is related to the problem of (top-$K$) view maintenance. The concept of detecting Hall of Fame events is related to general framework for event detection.

Research in the area of database mining is usually concerned with finding high support patterns (rules) in a given database [3], [4], [5], [6]. These patterns are often represented as frequent itemsets, association rules, causal rules, or mutual dependencies. Finding patterns (rules) in a database that do not have high support but are rather peculiar is addressed in [7], where the authors focus on finding peculiar data records based on the distance of their attributes from the majority of data items. Pattern mining in this case is done on previously retrieved peculiar data. Similar work has been done for graph data [8], where abnormal nodes are identified with their explanation being proposed automatically. Approaches [9], [10] extend this idea of pattern mining to data spread over multiple databases.

In event detection tasks, the goal is to detect noteworthy changes observed in dynamic data—for instance for early epidemic identification [11], [12] or topics detection in social network streams [2], [1]. Mining social network streams (or RSS feeds over blogs) could potentially be used to feed our underlying database with updates, in case of topics that are discussed there, such as soccer games, etc.

Exploring aggregated data based on different categorical constraints is the basic idea behind the data cube [13] technique for online analytical processing (OLAP) [14]. In the process of generating Hall of Fame queries, such (aggressive) pre-aggregations could be beneficial, but are deemed a large overhead in the maintenance task against database updates [14]. We materialize only the top-K results of suitable queries and use mechanisms to select a small subset of them as candidates to be updated.

Maintaining a set of Hall of Fame rankings over updates received to the underlying database is in its essence a task of matching updates against a set of registered queries. That is, identifying those materialized views (derived relations) in a database that are affected by an update [15]. Incrementally maintaining views that are affected by updates is considered, for instance, in [16]. Maintaining special case of views, namely top-$K$ views, is addressed in [17], [18] with the goal to render top-$K$ views self-maintainable, where updates are assumed to contain all necessary information with respect to the specific view.

The general concept behind continuously identifying events is also related to publish/subscribe approaches [19] where published documents are matched against user defined subscriptions, and has lately also been considered in top-$K$ query processing over data streams [20], [21], [22].

| Company | | | Person | | | Country | | StockMarket | |
|---|---|---|---|---|---|---|---|---|---|
| **c_id** | **c_name** | **c_countryid** | **p_id** | **p_name** | **p_countryid** | **co_countryid** | **co_name** | **s_companyid** | **s_value** |
| 0 | SAP | 0 | 0 | Bill Gates | 1 | 0 | Germany | 0 | 15 |
| 1 | BMW | 0 | 1 | Amancio O. Gaona | 5 | 1 | USA | 1 | 11 |
| 2 | Mercedes | 0 | 2 | Ingvar Kamprad | 4 | 2 | Italy | 2 | 16 |
| 3 | Microsoft | 1 | 3 | Warren E. Buffet | 1 | 3 | Greece | 3 | 115 |
| 4 | Google | 1 | … | … | … | 4 | Sweden | 4 | 255 |
| 5 | Facebook | 1 | | | | 5 | Serbia | 5 | 95 |
| 6 | Fiat | 2 | | | | … | … | … | … |
| 7 | Aegean Airlines | 3 | | | | | | | |
| 8 | Superfast Ferries | 3 | | | | | | | |
| … | … | … | | | | | | | |

| Shareholder | | |
|---|---|---|
| **s_personid** | **s_companyid** | **s_amount** |
| 0 | 3 | 400 |
| 0 | 5 | 50 |
| 1 | 8 | 40 |
| … | … | … |

$\mathcal{E}$-**attributes:** c_name, p_name, co_name
$\mathcal{C}$-**attributes:** c_name, co_name
$\mathcal{N}$-**attributes:** s_value

Fig. 2. Illustrative Schema and Data Instances for the "Bloomberg Billionaire Index" example.

Apart from the general concept of matching queries and documents, the work on processing top-$K$ queries over data streams [20], [21], [22] is fundamentally different from our problem of detecting events in Halls of Fame. In these works, the goal is to match arriving full-fledged documents against a set of user defined queries, dealing with adaptations of top-K aggregation algorithms [23] and partially consider skylines [24] for result maintenance, as objects (mainly unstructured text documents) are considered of limited lifetime.

In this work, we use database triggers (cf., e.g., [25]) as one way to implement the required event detection functionality, directly inside a Postgresql database.

When the schema of the database contains multiple tables without foreign key relations specified, approaches for attribute detection, such as the recent approach by Zhang et al. in [26], can be used to identify columns containing the same or similar attribute data—to find columns that can be used in the process of assembling Hall of Fame criteria.

## III. MODEL AND PROBLEM STATEMENT

We consider a set $\mathcal{R}$ of relations $R_i$ with schema $sch(R_i) = \{A_{i_1}, A_{i_2}, ...\}$ where the attributes $A_{i_j}$ contain entities, categories, and numerical values. We assume that an expert user is annotating the schema (Figure 2) to classify the attributes according to categories we define in the following. This is a reasonable assumption as it is a one-time process and also assumed to be a rather trivial task (for a person familiar with the database).

*Definition 1:* **Entity Attributes ($\mathcal{E}$-attributes):** An attribute $A_{i_j} \in sch(R_i)$ is said to be an entity attribute if the attribute instances represent real-world entities or concepts, such as persons, cities, or companies.

*Definition 2:* **Categorical Attribute ($\mathcal{C}$-attribute):** It is an attribute $A_{i_j} \in sch(R_i)$ the values of which are used to restrict the selected entities to a subset that has some specific property. For instance, in the example shown in Figure 1, country and company are marked to be a $\mathcal{C}$-attribute.

*Definition 3:* **Numerical Attribute ($\mathcal{N}$-attribute):** An attribute is called an $\mathcal{N}$-attribute if it can be used to compare entities based on their performance, that is, to rank entities and determine the outstanding (top) portion of them.

While $\mathcal{N}$-attributes are used as criteria to judge whether an entity can be a member of the Hall of Fame or not, $\mathcal{C}$-attributes restrict the focus of the decision to a specific group of entities.

Note that the sets of $\mathcal{E}$-attributes and $\mathcal{C}$-attributes do not have to be disjoint. On the contrary $\mathcal{N}$-attributes usually comprise a separate set of attributes as by nature they are inappropriate either as $\mathcal{E}$-attributes or as $\mathcal{C}$-attributes.

*Definition 4:* **Hall of Fame:** A ranked list of the best performing entities in a certain group is called Hall of Fame. It is defined by one $\mathcal{E}$-attribute, restricted to a certain group by a set of $\mathcal{C}$-attributes, and ranked according to one $\mathcal{N}$-attribute. A set of relations $\mathcal{R}$ can have many different Halls of Fame, depending on the attributes of the relations.

*Definition 5:* **Event:** An event is a change in a Hall of Fame caused by updates to the underlying database. An event has a number of characteristics. Some of them are static and depend on the query structure and others are dynamic and, thus, different in each update.

In this work we use the notion of Hall of Fame and Hall of Fame ranking interchangeably. We will see below that a Hall of Fame is generated by a query, hence, we sometimes also say Hall of Fame query to explicitly refer to its SQL/query character.

### A. Problem Statement

Given a set of relations $\mathcal{R} = \{R_1, R_2, ... R_n\}$ with user-defined annotations describing $\mathcal{E}$-attributes, $\mathcal{C}$-attributes, and $\mathcal{N}$-attributes the task is to:

1. **generate** queries that reflect Hall of Fame rankings
2. **maintain** the results of these queries against changes to the underlying database
3. **rank** identified changes (i.e., events) according to a model that reflects user-perceived interest

Note that generating Hall of Fame rankings involves building joins to connect $\mathcal{E}$-attributes, $\mathcal{N}$-attributes, and $\mathcal{C}$-attributes across different relations.

```
SELECT 𝓔-attribute, aggregate(𝓝-attribute)
FROM dataTable
WHERE predicate
GROUP BY 𝓔-attribute
ORDER BY aggregate(𝓝-attribute) ranking
LIMIT K
```

Fig. 3.  The Hall of Fame Query Template

## IV. Framework

The notion of a Hall of Fame naturally implies that only few entities are members of it—the top ones. This notion is directly reflected in the concept of top-K queries that restrict the query result to the best/top K results according to a given ranking criterion. In this work, we represent each Hall of Fame with a SQL style top-K query.

As updates are performed, Hall of Fame rankings change—entities enter or leave the Hall of Fame or just change ranking positions inside the top-K. Each identified change is considered for a certain amount of time—not only once. This time interval is application-dependent and would adhere to native time ranges of the scenario, like a day for stock markets or entire seasons/tournaments in sports. The idea here is to allow small (consecutive) rank improvements to be considered together, in this time interval; not to loose tiny improvements that essentially constitute a grand improvement. Identified changes are ranked and reported, as described in Section V.

### A. Hall of Fame Query Structure

Figure 3 shows the generic SQL query template we are using in this work. In the **FROM** clause of the query, the *dataTable* refers to one or multiple tables—connected through joins—that determine the schema and data available to this Hall of Fame. The *predicate* is a selection of a subset of data items corresponding to a certain aspect, and *ranking* determines the ordering (i.e., ascending or descending). By using the **LIMIT** clause (also called "fetch first"), the number of query results is restricted to at most K tuples, which constitutes the size of the Hall of Fame ranking.

*Predicate:*
A predicate used in the **WHERE** clause of a Hall of Fame query is a combination of **𝓒-attribute bindings**, **constant comparisons**, and **inter-attribute constraints**: For instance, $age > 21$ is a comparison with a constant, $assists \leq steals$ is an inter-attribute constraint, and the constraint *country = 'France'* or *city='New York'* is a 𝓒-attribute binding. In the rest of the paper, when no distinction is needed, we simply call them constraints.

The 𝓒-attribute bindings are generated using the actual data inside the relation. If there are two or more 𝓒-attributes in the predicate, we use all *possible* combinations of bindings —not blindly all combinations (i.e., the Cartesian product). For instance, consider the case of *country* and *city* that are both categorical attributes; we use the valid pairs of countries and cities that are materialized in the data, avoiding upfront

```
<comp> ::= > | < | = | ≠ | ≤ | ≥
<𝓒-attribute-bindings> ::= c ∈ 𝓒-attributes = <const>
<const> ::= "data that occurs in the corresponding column"
<inter-att-comparison> ::= c_1 ∈ sch(R_i) <comp> c_2 ∈ sch(R_j)
<anyconst> ::= "any constant with suitable type"
<const-comparison> ::= c ∈ 𝓒-attributes <comp> <anyconst>
<predicate> ::= <𝓒-attribute-binding> | <const-comparison>
              | <inter-att-comparison>
              | <predicate> ∧ <predicate> | true
```

Fig. 4.  BNF-style description of the aspect group constraints allowed in this work. Most importantly, we allow only conjunctions of predicates and only the equal (=) operator for attribute bindings.

bindings of the form *country='Germany'* and *city='Chicago'* that would lead to an empty query. While these 𝓒-attribute bindings are generated automatically, constant comparisons and inter-attribute constraints are specified by the users, while annotating the schema.

We limit the maximum number of constraints that can be used in a Hall of Fame predicate to a predefined threshold $cNum$—say four or five. This causes hardly any limitations in practice as a change to a Hall of Fame that uses many constraint conditions is very specific and in general hard to comprehend/appreciate for users.

Figure 4 contains a BNF (Backus-Naur Form) -style description of the possibilities to compose predicates to be used in Hall of Fame queries.

This means, we consider only predicates consisting of conjunctions of atomic predicates. While for the 𝓒-attribute bindings we use only *equal to* (=), for constant and inter-attribute comparisons we allow more comparison operators.

The above limitations on the complexity of Hall of Fame queries are currently inevitable to avoid an intractable number of generated queries that also need to be maintained.

*Ranking Entities:*
While the predicate confines a Hall of Fame to a specific subset of all entities, the order among entities is determined by numeric attributes. When expert users annotate the database schema, they do not only point to numeric attributes but also define ranking criteria; in a per-column manner irrespective of any later use in a specific Hall of Fame query. We show an example of such annotations in Table I (Section VII), where we introduce the evaluation dataset. The given annotations specify the columns that represent numeric attributes (of interest), the aggregation function, and the ranking order (i.e., ascending/descending or both).

### B. Query Generation

We create one SQL query per Hall of Fame following the structure in Figure 3. Considering all different possibilities for the 𝓔-attributes, 𝓒-attributes, and ranking criteria quickly results in a large number of combinations. In case the attributes in the query are from different tables, we enumerate join combinations that are required to connect tables that hold attributes needed to build Halls of Fame. A query is created for each join, $\mathcal{E}$-attributes, $\mathcal{C}$-attributes, and ranking criteria combination if none of the following two limitation for Hall of Fame queries are violated: (i) As mentioned previously, the number of $\mathcal{C}$-attributes that can be combined in a Hall of Fame predicate can be at most $cNum$. (ii) Additionally, a Hall of Fame query cannot have more that $jNum$ joins.

In this paper, all syntactically valid (i.e., joinable) and required join combinations are used. However, in case of a large number of tables, this set of join combinations could be controlled by an expert user. This would be a one-time process executed at startup and is not the scope of this paper.

We start the query generation by creating all possible constraint combinations, see Algorithm 1 for details. A constraints combination is possible if it is a conjunction of at most three constraints which are joined with each other with at most $jNum$ joins. For each such combination we query all distinct values. These values are used to create the attribute bindings of a Hall of Fame predicate. These predicates are then divided in groups according to the number of constraints they use. We start using the group with the least constraints and continue to those with more constraints.

For each constraints combination we check each $\mathcal{E}$-attribute. If the $\mathcal{C}$-attribute combination and the $\mathcal{E}$-attribute can be combined using no more than $jNum$ joins then we check all predicates that use this $\mathcal{C}$-attributes combination. For each predicate, we check whether the Hall of Fame that is produced *has at least K results*. Only in this case we create the Hall of Fame query by adding now a ranking criterion. Again, the Hall of Fame query is created only if the addition of the ranking criterion does not result in the need of more than $jNum$ joins. A pseudocode of this procedure is shown in Algorithm 2.

As the number of possible Hall of Fame queries is big, the above procedure ensures an early pruning of the possible combinations. For example, by querying the database for all distinct values that satisfy each constraints combination we avoid creating unnecessary queries, which would anyway be discarded later as empty queries. Further, we create predicates

**function** *get_combinations()*
  *combinations.add(empty_set)*
  **for each** *constraint*
  | **for each** *c* **in** *combinations*
  | | *c_new = c.copy.add(c_attr)*
  | | **if** *(c_new.size < cNum* **and** *joinable(c_new)*
  | |   **and** *join_size(c_new) < jNum)*
  | |   *combinations.add(c_new)*
  | | **end**
  | **end**
  **end**
  **return** *combinations*
**end**

Alg. 1: Create $\mathcal{C}$-attributes Combinations to be used in Hall of Fame query predicates.

*pruned = empty_set*
*combinations = get_combinations()*

**function** *prune_away(e_attr, comb, rank = 'any')*
  **for each** *c* **in** *combinations*
  | **if** *comb.is_subset_of(c)*
  |   *pruned.add(e_attr, comb, rank)*
  | **end**
  **end**
**end**

**function** *create_queries()*
  *queries = empty_set*
  **for each** *e_attr*
  | **for each** *comb* **in** *combinations*
  | | **if** *pruned.contains(e_attr, comb)* **break**
  | | **if** *join_size(e_attr, comb) > jNum*
  | | | *prune_away(e_attr, comb)*
  | | | **break**
  | | **end**
  | | *values = select_distinct(e_attr, comb)*
  | | **for each** *inst* **in** *values*
  | | | *results = exec_query(e_attr, comb, inst)*
  | | | **if** *results.size < K* **break**
  | | | **for each** *rank* **in** *ranking_criterion*
  | | | | **if** *pruned.contains(e_attr, comb, rank)* **break**
  | | | | **if** *join_size(e_attr, comb, rank) > jNum*
  | | | | | *prune_away(e_attr, comb, rank)*
  | | | | | **break**
  | | | | **end**
  | | | | *query = new_query(e_attr, comb, rank, inst)*
  | | | | *queries.add(query)*
  | | | **end**
  | | **end**
  | **end**
  **end**
  **return** *queries*
**end**

Alg. 2: Create Hall of Fame Queries

starting from those that contain zero or one constraint and consecutively enlarge such predicates by adding more constraints. This incremental process allows avoiding combinations of $\mathcal{E}$-attributes and constraints that produce Hall of Fame queries with too few results. This is easily doable since we allow only predicates that consist of conjunctions of atomic predicates: if a predicate A is not satisfiable, so is the predicate that contains A.

## V. EVENT RANKING

Detected changes to Hall of Fame rankings can be processed through higher-level applications, e.g., by narrative science approaches that create full-fledged newspaper articles or can be directly consumed by users without such pre-processing.

| pos. | p_name | sum(m_value) |
|---|---|---|
| 1. | William Henry Gates III | 210 |
| 2. | Warren E. Buffet | 210 |
| 3. | Amancio Ortega Gaona | 204 |

(Ranking before Update)

**UPDATE** StockMarket
**SET** s_value=s_value+10
**WHERE** s_companyid=8

(Update)

| pos. | p_name | sum(m_value) | |
|---|---|---|---|
| 1. | Amancio Ortega Gaona | 214 | ↗ +2 |
| 2. | William Henry Gates III | 210 | ↘ −1 |
| 3. | Warren E. Buffet | 210 | ↘ −1 |

(Altered Ranking)

Fig. 5. Illustration of an Update causing a Hall of Fame event.

When multiple events co-occur, a ranking schema is needed to order events in a way that reflects a user perceived notion of interestingness. This situation can occur due to large number of Hall of Fame queries or frequent updates, or when reflecting changes after a longer time period—such as after a completed season in soccer.

The ranking criterion to be used has to consider the specific characteristics of the query (e.g., its selectivity) and the properties of the change itself (e.g., its magnitude).

To combine several ingredients, we opt for using lexicographic tradeoffs [27]; they provide an elegant way to reconcile scores which per se can not be easily aggregated. Lexicographic orders with tradeoffs put emphasis on some aspects (scores) of the entities while using less important aspects to break the ties of the more important ones. This feature is important as we want to emphasize the competition implied by the selectivity of the query followed by the dynamic properties of the event, and finally the characteristics of the constraints used.

Plain lexicographical ordering is a very common technique known from (printed) dictionaries or encyclopedia: Two strings are compared character-by-character, usually from left to right. The first position at which the characters differ decides the order of the two strings. This procedure can be generalized to arrays (strings) of numerical values instead of characters.

However, when applied to numerical values, especially non-integer domains, ties are highly unlikely to occur. In this case, the first position imposes the order of items. To circumvent this, lexicographic tradeoffs introduce variations of the comparison at the individual positions. Given two values $v$ and $u$, one could define $v <_{lt} u$ to be true iff $v$ is not only smaller than $u$ but "considerably smaller", and the same for $v >_{lt} u$ iff $v$ is "considerably larger". Consider for example the following comparison function $v <_{lt} u$ only if $2*v \leq u$ and $v >_{lt} u$ if $v > 2*u$. Then $(7, 3, 6)$ is larger than $(3, 8, 4)$, since by comparing the values in the first position we have $7 > 2*3$. On the other hand, $(7, 2, 6)$ is smaller than $(5, 6, 2)$, since for the values in the second position we have $2*2 < 6$ (the values in the first position report the two sequences to be equal). Hence, from left to right, the first "not-equal" value decides the ordering of the sequence.

We first introduce in details the individual ingredients of our ranking schema and then show how to combine them.

### A. Dynamic Characteristics

Hall of Fame events are characterized by a distortion of its entity ranking. Figure 5 shows such an event for *Amancio*

**entity $e_1$:**
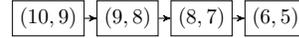

**entity $e_2$:**
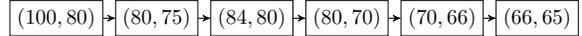

Fig. 6. Information at hand for computing the dynamic score component

*Ortega Gaona* who moved up from position 3 to position 1 in our toy example of a billionaires index. Every time an entity improves from rank $r$ to rank $r'$ (i.e., $r > r'$) the maintenance component forwards this information to the ranking framework. Figure 6 illustrates the available information for two entities, $e_1$, and $e_2$, with respect to one specific Hall of Fame. We see pairs of ranks that indicate the changes (only decreasing ranks).

As mentioned in Section IV, this bookkeeping is done only for a limited amount of time and takes care that individual rank improvements should be considered together. For instance, at the end of a season, rank improvements over the entire season should be considered, not only the last improvement (which might be negligible itself).

To compute a score describing the quality of the jump (from rank $r$ to rank $r'$), for each entity, the most recent time point (i.e., now) of each such chain of pairs is considered. The individual scores for each pair are summed up to the point the sequence would continue with rank that is lower than the currently highest rank. In our example, illustrated in Figure 6, for the entity $e_2$ we consider the improvement from rank 84 to 65, and not from 100 to 65. The earlier jump from 100 to 75 had the chance of being reported at an earlier time point.

For each individual improvement from rank $r$ to rank $r'$, the size of the improvement (i.e., $r-r'$) is considered to resemble the importance of the improvement. Pairs of the form $(r-r')$, for the same entity and Hall of Fame query, are then aggregated: instead of purely summing up these pairs, the impact of improvements at lower ranks is decreased. This resembles the concept of Discounted Cumulative Gain (DCG) [28] used in information retrieval to model the user perceived information quality of a ranking of quality assessments.

$$rs(\{(r_i, r'_i) | i \in \mathbb{N}\}) := \sum_{r'_i \leq b} (r_i - r'_i) + \sum_{r'_i > b} \frac{(r_i - r'_i)}{log_b(r'_i)}$$

This score gives weight 1 to all rank improvements above rank $b$, and punishes lower ranks with a weight of $1/log_b(.)$

This dynamic score results in values between $1/log_b(K)$ (for the smallest noticeable rank improvement from rank $K+1$ to rank $K$) and $K$ (for the biggest noticeable improvement from rank $K+1$ to rank 1). We use these two values with min-max normalization to normalize the final dynamic score values to $[0, 1]$ range.

*B. Static Characteristics*

Consider the following query that describes the top-20 male players in the North American Basketball Association (NBA) at the age of 24, ranked according to the total number of points scored.

**SELECT** name, **SUM**(points)
**FROM** players
**WHERE** gender='male' **AND** league='NBA' **AND** age=24
**GROUP BY** name
**ORDER BY sum**(points) **DESC**
**LIMIT** 20

The aspect group constraint selects a subset of the entire table *players* using constraints on the attributes *team*, *gender* and *age*. There might be also other queries that use constraints on *nationality* or *height* as selection criteria. In order to rank different queries, the ability of the aspect group constraint to create useful groups is taken into consideration. This is done by computing the information entropy [29].

Complementary to the entropy of the aspect group, the *selectivity* of the aspect group constraint is the second ingredient of the static score of a Hall of Fame query and is simply defined as the selectivity of the where clause of the Hall of Fame generating query.

*1) Entropy:* Each Hall of Fame predicate consists of a set of attributes used in $\mathcal{C}$-attributes bindings (e.g., team='Phoenix'). All these attributes restrict the focus of the query to a certain subset of the database tuples. We now assess how informative the groupings, created from the combination of attributes in the where clause, are.

To achieve this, the table is projected to the columns that are used in the Hall of Fame predicate. For instance, in the case of a query with where clause "*team='Phoenix' AND year=1995 AND league='NBA'* ", the data is projected to the columns *team, year*, and *league*. We calculate the informativeness of the grouping by calculating the entropy [30] of the projected data. We use entropy in a similar way used in the data mining component of MS SQL Server 2008 R2 [31] to find meaningful categorical attributes.

This is done as follows: Suppose there is a table $T$ which has the attributes $attr_1, attr_2, \ldots, attr_n$. We inspect all possible instantiations, $I_i$, of the set $(attr_1, attr_2, \ldots, attr_n)$ using the data values found in the table. For each instantiation $I_i$ we count the number of tuples that satisfy it and then divide with the total number of tuples in the table $T$. This is the probability of an instantiation $I_i$. We then compute the entropy of the table using Shannon Entropy ([29]).

For example, assume we have the following projection

| team | year | league |
|---|---|---|
| Phoenix | 2010 | NBA |
| Boston | 2011 | NBA |
| Phoenix | 2010 | NBA |
| San Antonio Spurs | 1972 | ABA |
| Phoenix | 2010 | NBA |

The different instantiations created from the above projection are:

$I_1$ = (Phoenix, 2010, NBA)
$I_2$ = (Boston, 2011, NBA)
$I_3$ = (San Antonio Spurs, 1972, ABA)

We can now compute the probabilities for these instantiations as

$$P(I_1) = 3/5$$
$$P(I_2) = 1/5$$
$$P(I_3) = 1/5$$

allowing us to calculate the final entropy.

*2) Selectivity:* The entropy of the Hall of Fame predicate is a general measure that applies to all queries having the same combination of attributes in the where clause.

However, we would like to assess not only the general quality of the predicate but also the quality of the specific instantiation that is used in a query.

The idea is, the smaller the fraction of the table that qualifies for a query result is, the less competitive the ranking is and, thus, the lower the score of such a query is. Or in other words, the higher the selectivity of the query the more interesting it is. This selectivity of the query can be easily computed as the selectivity of the "where clause", i.e., the selectivity of the predicate.

*C. Putting it All Together*

The above reasoning created three different components to consider for an overall ranking: selectivity, dynamic score, and entropy.

As mentioned, we use lexicographic tradeoffs [27], which sort sequences of values based on primary, secondary, and so on, criteria, that correspond to entries in the sequence (say from left to right). In our case, the selectivity of the underlying query, representing the competition between the entities, is the decisive factor and is used as primary criteria, followed by the dynamic score and the entropy in the end.

Taking the plain values for these three individual measures will not lead to ties at the specific sequence positions and, hence, only the first criterion will be used to sort.

To avoid this, for each one of the primary and secondary criteria, the following comparison function is used: $u <_{lt} v$ *iff* $(u + \frac{1}{2n}) <_{lt} v$ where $n$ is the number of coarse groups that can be identified in the data, with respect to the ranking criterion. It is important to note that scores $u, v$ are considered as normalized in the range $[0, 1]$ with 0 denoting the lowest possible and 1 the largest possible score. This creates a range of size $1/n$ of "equal" values around each original value (Figure 7), which allows values to be considered equal even if their original numeric values differ (slightly).

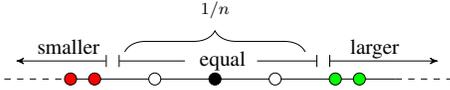

Fig. 7. Ranges in Lexicographic Tradeoffs

Orientation for a reasonable choice of the size of these ranges could potentially be found in the number of semantically meaningful categorical constraints, although there is no universally true rule for this. Intuitively, there are really big Halls of Fame featured by their high selectivity, like the top-20 richest people in the entire world. The large ones, like the top-20 richest people in Europe. The medium ones that have low selectivity, like the top-20 richest people in a country. And the small ones that have very low selectivity, like the top-20 richest people in a city.

Similar to the selectivity of a Hall of Fame query, dynamic scores can also be thought of as being decisive only if they notably differ. For instance, a change at rank 5 is clearly more important than a change at rank 50, but it is questionable if a change at rank 6 is really less important to the one at rank 5. For the last ingredient, i.e., the entropy based measures, we do not need to define a new comparison function as it has the lowest priority in the lexicographic sorting.

## VI. EVENT DETECTION

Detecting Hall of Fame events requires monitoring Hall of Fame rankings for changes. As we are interested in the positions of entities inside the rankings, a change is considered only if the ranking changes, not if entities obtain updates to their numeric score (aggregate)—but the ranking remains unchanged. A naïve approach to this monitoring problem would execute all queries that underlie the Hall of Fame rankings and check if their results have changed. Instead, we make use of techniques from the area of materialized view maintenance. In this, there are two types of optimization techniques. First, we can use information from the update statement itself to determine the queries that might be affected by the update and only refresh those. See [15] for a formal analysis of the task. Second, after an affected Hall of Fame ranking is identified one can optimize the actual re-evaluation. This task is also known as view maintenance and has been addressed for top-$K$ queries [17], [18]. Having tens of thousands of query, the main focus of the event detection is to identify possibly affected ones, rather than to optimize their re-evaluation.

We employ *two filtering techniques* that tell if a Hall of Fame cannot be affected by a given update—and, hence, does not have to be re-evaluated. Both techniques are precise in the sense that they do not produce false-positives (i.e., the filter says the ranking is not affected but actually it is) but they do produce false-negatives (i.e., the filter says the ranking might be affected but it is not) i.e., passing the filter is a necessary condition to lead to a Hall of Fame event, but it is not sufficient to tell if the ranking is affected.

1. **Column-Based Filtering** resembles a satisfiability check that uses the overlap between the attributes used in the Hall of Fame queries and the update statement. Clearly, a query can not be affected by an update if the intersection between the columns being updated and the columns accessed by the query is empty.

The column-based filtering is executed solely on the Hall of Fame specification and the update statement. There are, however, many Halls of Fame that cannot be pruned based on this filter. To further limit unnecessary query re-evaluations, we use another level of filtering.

2. **Row-Based Filtering** tells if a query can not be affected by an update by inspecting the intersection between the updated rows and the rows accessed by the query.

The queries that pass these two filters are re-evaluated and checked for potential changes to their rankings—which would result in events.

### A. Implementations

*1) Database Triggers:* When designing the implementation of such a maintenance task, a first choice is to use database triggers and materialized views, since the database is already used to store the data itself—such an implementation is elegant and intuitive: all processing is done in the database, hence, there is no additional layer. However, pushing the maintenance task in the database also limits the capabilities of the database to serve other applications. One additional feature of this approach is that identified events are stored directly inside the database—written to disk, hence, being persistent. This, of course, comes at the price of increased latency.

Identified events are appended to a special events table, that contains the score ingredients as introduced in Section V.

*2) Event Processing Middleware:* As an alternative implementation, we have implemented an event processing middleware using Java. Only the data tables themselves are still stored in a database—the Hall of Fame rankings are kept in main memory. The middleware intercepts all updates sent to the database and checks if Hall of Fame rankings are affected. By doing this, we alleviate any additional (event processing) work of the database but introduce additional communication cost between the database and the Java application—which, however, appears negligible as we run the middleware and the database on the same server. Further, all results are easily kept in main memory, which should makes the processing faster, but comes of course with the risk that after a system failure the monitoring state is lost—not the database content. Still, this is acceptable in most scenarios and, if not acceptable, can be circumvented, for example, by replication.

We implemented both approaches (cf., Figure 8) using Java 6 and Postgresql 9.1 as the database, with column- and row-based filtering in both implementations. In Postgresql, we use additional tables and triggers to store the Hall of Fame rankings and to check for modifications.

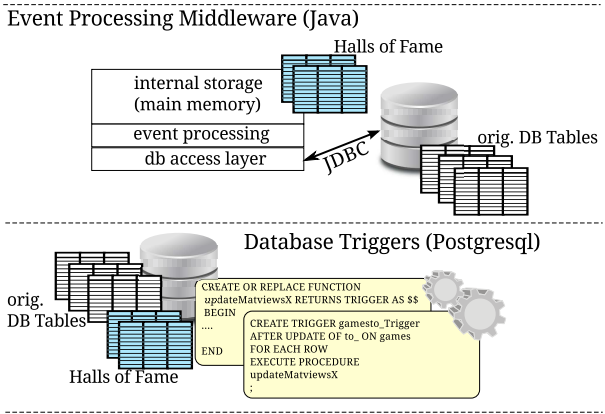

Fig. 8. The two implemented Event Detection Approaches: Two-Tier architecture of a Postgresql database and a Java middleware (top) and implemented inside Postgresql using Triggers

*3) Column-Based Filtering Implementation:* The implementations of the column-based filtering in Java and Postgres triggers is conceptually the same: an index is created such that for a given column name it contains all of the queries that depend on that column. This index is created once all the Hall of Fame queries are generated. While in Java a HashMap is used, the trigger implementation uses one stored procedure per column name, with a predefined list of queries which might be affected. This is done for efficiency reasons, such that querying an additional table containing the column index is avoided.

*4) Row-Based Filtering Implementation:* The row-based filtering requires a selection of all rows that are going to be updated by the update statement. This is achieved in the trigger-based implementation using *row triggers*—a stored procedure is invoked for each row that is being updated. When the queries depend on multiple tables, rows that are updated are progressively joined with other tables. For each update and all queries that pass the column filtering a check is performed, based on the query selection criterion—whether this row is selected by the query. If the query selects any of the updated rows it is potentially affected.

In the Java middleware, we avoid progressively joining updated rows by creating a set of *select* queries covering all join paths. Creating these queries starts from the updated table and follows foreign keys to join with this table. In case there are two or more joins paths between two tables, two or more copies of the query are created. Depending on the structure of the Hall of Fame query the selected rows by one of corresponding selection queries are used. The pseudo code for creating selection queries is shown in Figure 3, where **join** method joins a path with an existing query if they are compatible. It is important to note that this query creation is done only once at the startup of the system, when the database schema is registered.

## VII. EXPERIMENTS

All experiments are conducted on a Linux machine (Debian 5.0.9 64bit, kernel 2.6.32.41.1) with two Intel Xeon W3530

```
function selection_queries(updated_table)
  tables = all_tables
  visited = empty_set
  queries = new set(empty_query)

  for each t in tables
  | if not visited.contains(t)
  | | paths = all_join_paths(updated_table, t)
  | | for each q in queries
  | | | q.join(paths.first)
  | | end
  | | for each path in paths.rest
  | | | for each q in queries
  | | | | new_q = q.copy()
  | | | | new_q.join(path)
  | | | | queries.add(new_q)
  | | | end
  | | | visited.add(path.tables)
  | | end
  | | visited.add(paths.first.tables)
  | end
  end
end
```

Alg. 3: Create row selection queries

CPUs (2.8GHz, 8MB cache, 4 cores (8 threads)). The machine has 48GB of main memory and a RAID-5 disk (4x1TB Western Digital WD1002FBYS-1 7.2K RPM, SATA).

### A. Dataset

For the dataset, we use publicly available basketball statistics obtained from databasebasketball.com [32]. We focus on the regular season statistics, which capture the performance of roughly 4000 players in the last 65 years. In this table, each player is uniquely identified by the player_id containing statistics for one year, per team. The latter is expressed by a team id which refers to a second table that contains 121 basketball teams with attributes team id, team name, and league. It is possible for players to switch teams in an active season (year). Attributes of the season statistics table describe the information about minutes played, points, points per game, total rebounds, free throws, etc.

We use the *player* and *team* columns as $\mathcal{E}$-attributes and believe that this is a reasonable choice for this scenario. To specify the ranking attributes, not only the column names but also the aggregation function and the "order by" direction (descending, ascending, or both) are specified (cf., Table I). The attributes *league*, *team*, and *age* are used as $\mathcal{C}$-attributes, with two additional categorical constraints specified: *steals>turnovers* and *offensive rebounds>defensive rebounds*. We created B+ indices for each single categorical attributes.

As the dataset reports only per-season statistics it lacks "live" data, that is, statistics that are updated after every game (or after every notable event, e.g., a scored point), we create updates based on the original data summaries.

| column name | aggr. function | order |
|---|---|---|
| turnovers | sum | ascending |
| rebounds | sum | descending |
| assists | sum | descending |
| field goals percentage | avg | both |
| points per game | avg | descending |
| points | sum | descending |
| three-points per game | sum | descending |
| three throws made | sum | descending |

TABLE I

SAMPLE USER GENERATED DESCRIPTION OF HOW TO USE NUMERIC ATTRIBUTES. USED IN THE EXPERIMENTAL EVALUATION.

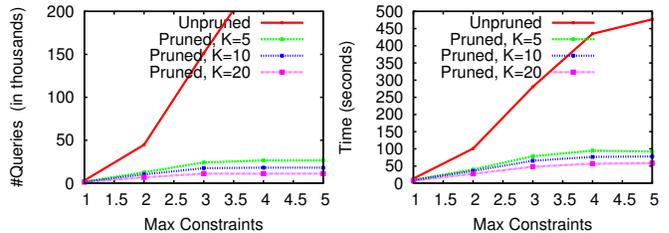

Fig. 9. Query Generation Process: Total number of generated queries (left) and total wall-clock time in seconds (right)

For each data tuple represented by a triple (player, team, year) and for each ranking attribute of interest (cf., Table I), we create 10 update statements where the intermediate values are calculated differently depending on whether summation or averaging is used as the aggregation function. The last update value always corresponds to the actual value in the database. The generated updates are spread equidistantly over time. To keep the evaluation tractable, we used the years 2005 to 2011, and generated this way a total of 275,580 updates.

For the summation, we model the values as steadily growing. We generate values using a Normal distribution (with $\mu = 0.5$, $\sigma = 0.2$) until we have 9 values in range $(0, 1)$. These values are sorted in ascending order. Finally, the value of the $i^{th}$ update equals to the $i^{th}$ generated value times the final (actual) value.

For the averaging, we model the values as slightly fluctuating around the final (actual) value. To generate this, we use a Normal distribution ($\mu = 0$, $\sigma = 0.1$) and generate 9 values in range $(-1, 1)$. The values are are not sorted as in the previous case. Again, the value of the $i^{th}$ update equals to the $i^{th}$ generated value times the final (actual) value.

*B. Query Generation*

The number of generated queries for 3 different values for the Hall of Fame ranking size $K$ is shown in Figure 9 (left). The y-axis represents the number of generated queries while the x-axis shows the maximum number of constraints used in a query. The curve for *unpruned* represent the total number of queries that would be generated in case no pruning of empty queries or queries with less than $K$ results is applied. As expected, this number grows rapidly, up to roughly 1 million queries. In the opposite case, when pruning is applied, very many queries can be removed. Again, the number of queries grows with the number of constraints used in the queries, but saturates at some point as newly generated queries with too many constraints have small (or empty) result sets and are discarded. For the same reason, increasing $K$, decreases the number of generated queries.

During the query generation, most time is spent in executing queries to obtain the initial results and to compute the static characteristics—used in the event ranking later-on—of the generated queries. The wall-clock time for the query generation is shown in Figure 9 (right). We compare the time needed to execute all generated queries without integrated pruning (the curve for *unpruned*) to the approach that eliminates queries from consideration that can be identified without executing them (cf., Section IV-B). The results are as expected: early pruning can greatly reduce the overall wall-clock time for the entire query generating process. The total wall-clock time for all early pruning cases ($K = 5$, 10, or 20) is below 100 seconds, which is well acceptable, given that that the query generation is a one-time process only.

*C. Ranking Quality*

We conducted a user study to evaluate the quality of the proposed event ranking. We present to users 9 rankings of events. Each ranking reflects the Hall of Fame events within the last 1000 updates, which corresponds roughly to the time period in which all teams have played one game. Considering that generated updates are equidistant in time, we maintain a windows of events corresponding to the last 1000 updates as well. We used queries generated with $K = 20$ and maximum of up to three constraints. Using $K = 20$ also influences the parameter $b$ for dynamic characteristics score, which we set to 5. To keep the assessment task manageable, for each user in a couple of minutes, each ranking consists of 3 events: the one put by our algorithm at rank one, rank five, and rank ten. For each such event, we created (manually) a natural language statement, such as *"Team Philadelphia advanced from position 19 to position 17 in the NBA top field game points scored list."*

We randomly order these sentences and ask users to order them using a scale from 1 to 3, also allowing ties. We asked 11 colleagues at our university to take part in the study. The summarized results are shown in Table II. We can clearly see a good trend of diagonal elements being the largest, as the perfect ranking would produce completely diagonal matrix.

For each ranking, we compute how well the assigned ratings of a user adhere to our proposed ranking. We inspect the following pairs of reported ranks: $(1, 5)$, $(1, 10)$, $(5, 10)$. The ideal user rating should give assessments that are coherent with those pairs. For instance, an ordering of ranked events $(5,1,10)$ would ideally be judged by the users with scores $(2,1,3)$. This fit is computed for each above pair $(a,b)$ using a value of 1 if $r(a) > r(b)$, a value of $-1$ if $r(b) < r(a)$ and 0 otherwise, where r(.) denotes the user's rating. We sum up this scores over all pairs, over all event rankings and over all users. and achieve 66 as our final result. Knowing the user study results we can calculate this result for the best possible

|  | **Assigned Ratings by Users** | | |
|---|---|---|---|
|  | Rating 1 | Rating 2 | Rating 3 |
| Event at Rank 1 | **49** | 34 | 16 |
| Event at Rank 5 | 24 | **52** | 23 |
| Event at Rank 10 | 29 | 25 | **45** |

TABLE II

RAW RESULTS OF OUR USER STUDY

| #Constraints | Queries Total | Queries Executed | Changes Detected | Runtime (ms) Triggers | Java |
|---|---|---|---|---|---|
| 1 | 1710 | 10.44 | 0.03 | 239.52 | 127.84 |
| 2 | 10260 | 20.11 | 0.51 | 774.29 | 185.93 |
| 3 | 17540 | 24.23 | 1.00 | 1427.82 | 213.25 |

TABLE III

EXECUTED QUERIES PER UPDATE AND EXECUTION TIME, $K = 10$

| #Constraints | Queries Total | Queries Executed | Changes Detected | Runtime (ms) Triggers | Java |
|---|---|---|---|---|---|
| 1 | 1290 | 10.43 | 0.12 | 234.98 | 130.05 |
| 2 | 6900 | 19.67 | 1.03 | 700.37 | 178.99 |
| 3 | 11120 | 23.26 | 1.82 | 780.26 | 207.67 |

TABLE IV

EXECUTED QUERIES PER UPDATE AND EXECUTION TIME, $K = 20$

ranking algorithm and the worst one, resulting in score $148$ for the best and $-148$ for the worst. Results for the best (worst) possible ranking must be calculated based on the user study results as the disagreement between users must be taken into account. We see that proposed ranking performs very well satisfying $72.39$ % ($\frac{66+148+1}{148+148+1}$) of users judgments.

To show that these numbers are statistically significant, we apply Fisher's randomization test (also called permutation test) (cf., [33]), where the null hypothesis claims that our ranking algorithm (scheme) would place events simply in a random order. The significance test showed that we can reject the null hypothesis with significance level of $3\%$ ($p = 0.0203$).

### D. Event Detection

To evaluate the event detection performance, we use $5,000$ updates and measured the number of executed queries per update, the number of detected changes, and the time needed to process each update for both Java and trigger-based implementation. The maximum number of constraints is varied from 1 to 3.

Table III and Table IV report on the total number of generated Hall of Fame queries, the average number of executed queries per update (i.e., queries that pass both column- and row-based filtering steps), the average number of queries per update for which the ranking changed, and the execution time per update in milliseconds for both implementations. Table III contains the measurements when queries are generated with $K = 10$, while Table IV reports on the measures with queries generated with $K = 20$.

Although there is a large number of queries generated and indexed, we can see that by using the filtering techniques we can already decrease the number of executed updates to less than one percent. This percentage drops significantly with the increase of the maximum number of constraints allowed in the queries. This shows that we can safely increase the maximum number of constraints while still being able to detect changes efficiently.

We see that the wall-clock time (in milliseconds) for both implementations is quiet similar for the simple case of at most one constraint per query. However, the runtime increases a lot faster for the implementation with triggers, resulting in a runtime up to $1.4$ seconds per update. We believe that this stems from the fact that results of the operations on the triggers are always made persistent, while the results of the Java implementation are always held in memory without writing them to disk. We observe that the wall-clock time behaves similarly to the measured number of executed queries, which suggests that most of the execution time is indeed spent executing queries. The wall-clock time when employing a Java main memory implementation is in all cases under $250ms$.

As expected, the number of detected changes per update increases as the number of queries grow, especially the queries with more constraints are more selective making an impact of the updates larger enough to change resulting ranking. It is interesting to see that increasing the value of $K$ (from 10 to 20) results in the smaller number of generated queries, as there are less queries with result size 20 and more, but the number of detected changes increases due to the more volatile nature of the rankings between position 10 and 20.

## VIII. CONCLUSION

Halls of Fame are by design intriguing constructs that represent the elite of a certain subset of entities. In this work, we studied the automatic identification of changes to Halls of Fame in a principled and general way. To the best of our knowledge, this problem has not been addressed before.

Our approach generates SQL queries that describe (each) a specific Hall of Fame. Such queries select the top-K entities that fulfill constraints put on categorical attributes, ranked according to specific numeric attributes. We discussed means to inspect only a small subset of cached query results for potential changes. The presented approach is very generic and can be applied to a wide range of scenarios. We presented the details of two fundamentally different implementations: (i) an approach based on database triggers and (ii) an event processing middleware implemented in Java.

We conducted a carefully designed experimental evaluation using real-world data obtained from a basketball statistics website. We showed that the number of generated queries saturates with growing number of possible categorical constraints. This is an important finding underpinning the applicability of our approach. Further, we saw that by using our filtering approach only a small fraction of queries need to be re-evaluated for incoming updates. Most importantly, the conducted user study showed that the proposed lexicographic-tradeoff–based ranking complies with a user-perceived notion of interestingness.

REFERENCES

[1] M. Mathioudakis and N. Koudas, "Twittermonitor: trend detection over the twitter stream," in *SIGMOD Conference*, 2010, pp. 1155–1158.
[2] F. Alvanaki, S. Michel, K. Ramamritham, and G. Weikum, "See what's enblogue - real-time emergent topic identification in social media," in *EDBT Conference*, 2012.
[3] R. Agrawal, T. Imielinski, and A. N. Swami, "Database mining: A performance perspective," *IEEE Trans. Knowl. Data Eng.*, vol. 5, no. 6, pp. 914–925, 1993.
[4] R. Agrawal and R. Srikant, "Fast algorithms for mining association rules in large databases," in *VLDB*, 1994, pp. 487–499.
[5] M. J. Zaki, "Efficiently mining frequent trees in a forest: Algorithms and applications," *IEEE Trans. Knowl. Data Eng.*, vol. 17, no. 8, pp. 1021–1035, 2005.
[6] M. A. Hasan, V. Chaoji, S. Salem, J. Besson, and M. J. Zaki, "Origami: Mining representative orthogonal graph patterns," in *ICDM*, 2007, pp. 153–162.
[7] N. Zhong, C. Liu, Y. Yao, M. Ohshima, M. Huang, and J. Huang, "Relational peculiarity oriented data mining," in *ICDM*, 2004, pp. 575–578.
[8] S.-D. Lin and H. Chalupsky, "Discovering and explaining abnormal nodes in semantic graphs," *IEEE Trans. Knowl. Data Eng.*, vol. 20, no. 8, pp. 1039–1052, 2008.
[9] S. Zhang, X. You, Z. Jin, and X. Wu, "Mining globally interesting patterns from multiple databases using kernel estimation," *Expert Syst. Appl.*, vol. 36, no. 8, pp. 10 863–10 869, 2009.
[10] S. Zhang, C. Zhang, and J. X. Yu, "An efficient strategy for mining exceptions in multi-databases," *Inf. Sci.*, vol. 165, no. 1-2, pp. 1–20, 2004.
[11] A. Cami, G. L. Wallstrom, A. L. Fowlkes, C. A. Panozzo, and W. R. Hogan, "Mining aggregates of over-the-counter products for syndromic surveillance," *Pattern Recogn. Lett.*, vol. 30, no. 3, pp. 255–266, Feb. 2009. [Online]. Available: http://dx.doi.org/10.1016/j.patrec.2008.09.008
[12] W.-K. Wong, A. W. Moore, G. F. Cooper, and M. M. Wagner, "Bayesian network anomaly pattern detection for disease outbreaks," in *ICML*, 2003, pp. 808–815.
[13] J. Gray, A. Bosworth, A. Layman, and H. Pirahesh, "Data cube: A relational aggregation operator generalizing group-by, cross-tab, and sub-totals," *Data Engineering, International Conference on*, vol. 0, p. 152, 1996.
[14] S. Chaudhuri and U. Dayal, "An overview of data warehousing and olap technology," *SIGMOD Record*, vol. 26, no. 1, pp. 65–74, 1997.
[15] J. A. Blakeley, N. Coburn, and P.-Å. Larson, "Updating derived relations: Detecting irrelevant and autonomously computable updates," *ACM Trans. Database Syst.*, vol. 14, no. 3, pp. 369–400, 1989.
[16] T. Griffin, L. Libkin, and H. Trickey, "An improved algorithm for the incremental recomputation of active relational expressions," *IEEE Trans. Knowl. Data Eng.*, vol. 9, no. 3, pp. 508–511, 1997.
[17] K. Yi, H. Yu, J. Yang, G. Xia, and Y. Chen, "Efficient maintenance of materialized top-k views," in *ICDE*, 2003, pp. 189–200.
[18] E. Baikousi and P. Vassiliadis, "Maintenance of top-$k$ materialized views," *Distributed and Parallel Databases*, vol. 27, no. 2, pp. 95–137, 2010.
[19] P. T. Eugster, P. Felber, R. Guerraoui, and A.-M. Kermarrec, "The many faces of publish/subscribe," *ACM Comput. Surv.*, vol. 35, no. 2, pp. 114–131, 2003.
[20] K. Mouratidis, S. Bakiras, and D. Papadias, "Continuous monitoring of top-k queries over sliding windows," in *SIGMOD Conference*, 2006, pp. 635–646.
[21] K. Mouratidis and H. Pang, "Efficient evaluation of continuous text search queries," *IEEE Trans. Knowl. Data Eng.*, vol. 23, no. 10, pp. 1469–1482, 2011.
[22] P. Haghani, S. Michel, and K. Aberer, "The gist of everything new: personalized top-k processing over web 2.0 streams," in *CIKM*, 2010, pp. 489–498.
[23] R. Fagin, A. Lotem, and M. Naor, "Optimal aggregation algorithms for middleware," *J. Comput. Syst. Sci.*, vol. 66, no. 4, pp. 614–656, 2003.
[24] S. Börzsönyi, D. Kossmann, and K. Stocker, "The skyline operator," in *ICDE*, 2001, pp. 421–430.
[25] N. W. Paton and O. Díaz, "Active database systems," *ACM Comput. Surv.*, vol. 31, no. 1, pp. 63–103, Mar. 1999. [Online]. Available: http://doi.acm.org/10.1145/311531.311623
[26] M. Zhang, M. Hadjieleftheriou, B. C. Ooi, C. M. Procopiuc, and D. Srivastava, "Automatic discovery of attributes in relational databases," in *SIGMOD Conference*, 2011, pp. 109–120.
[27] R. Luce, "Lexicographic tradeoff structures," *Theory and Decision*, vol. 9, pp. 187–193, 1978.
[28] K. Järvelin and J. Kekäläinen, "Cumulated gain-based evaluation of ir techniques," *ACM Trans. Inf. Syst.*, vol. 20, no. 4, pp. 422–446, 2002.
[29] C. Manning and H. Schutze, "Foundations of statistical natural language processing." MA, USA: MIT Press, 1999. [Online]. Available: /brokenurl#http://publication.wilsonwong.me/load.php?id=233281606
[30] E. Unger, L. Harn, and V. Kumar, "Entropy as a measure of database information," in *Sixth Annual Computer Security Applications Conference*, 1990, pp. 80–87.
[31] "Microsoft SQL Server 2008 R2: Feature selection in data mining. http://msdn.microsoft.com/en-us/library/ms175382.aspx."
[32] "databasebasketball.com - nba basketball statistics, draft, awards, and history http://www.databasebasketball.com/."
[33] M. D. Smucker, J. Allan, and B. Carterette, "A comparison of statistical significance tests for information retrieval evaluation," in *CIKM*, 2007, pp. 623–632.